\documentclass{iau} 
\usepackage{graphicx}
\usepackage{hyperref}

\def\aap{A\&A}                
\def\mnras{MNRAS}             
\def\apj{ApJ} 
\def\aj{AJ}        
\def\pasp{PASP}

\title[Binary interaction along the RGB] 
{Binary interaction along the RGB:\\
\Large
The Barium Star perspective}

\author[Escorza et al.]   
{Escorza, A.$^{1,2}$
,
Siess, L.$^{2,3}$,
Karinkuzhi, D.$^2$,
Boffin, H.M.J.$^{4}$,\\
Jorissen, A.$^2$ \and
Van Winckel, H.$^1$}

\affiliation{$^1$ Institute of Astronomy, KU Leuven, Celestijnenlaan 200D, B-3001 Leuven, Belgium \\ email: {\tt ana.escorza@kuleuven.be} \\[\affilskip]
$^2$ Institut d'Astronomie et d'Astrophysique, Universit\'e Libre de Bruxelles, Campus Plaine C.P. 226, Boulevard du Triomphe, B-1050 Bruxelles, Belgium \\[\affilskip]
$^3$ F.R.S.-FNRS, Belgium\\[\affilskip]
$^4$ ESO, Garching bei M\"unchen, Germany\\[\affilskip]
}

\pubyear{2018}
\volume{343}  
\jname{Why Galaxies Care About AGB Stars}
\editors{F. Kerschbaum, M. Groenewegen \& H. Olofsson}
\begin{document}

\maketitle

\begin{abstract}
Barium (Ba) stars form via mass-transfer in binary systems, and can subsequently interact with their white dwarf companion in a second stage of binary interaction. We used observations of main-sequence Ba systems as input for our evolutionary models, and try to reproduce the orbits of the Ba giants. We show that to explain short and sometimes eccentric orbits, additional interaction mechanisms are needed along the RGB.

\end{abstract}

\firstsection 
\section{Introduction}
Stars with extended convective envelopes, like red or asymptotic giant branch (RGB or AGB) stars, in binary systems can fill a substantial fraction of their Roche Lobe and exchange angular momentum and possibly mass with their companion. In this study, we considered a family of chemically peculiar stars known as barium (Ba) stars to investigate binary evolution along the RGB.

Ba stars
are main-sequence or giant stars formed due to binary interaction when a former AGB companion, which is now a dim white dwarf (WD), polluted them with heavy elements \cite[(McClure 1984)]{McClure84}. The interaction with the former AGB companion, i.e. the formation of the Ba star, is not well understood. However, we focused on the evolutionary link between dwarf and giant Ba stars which could be affected by a second phase of binary interaction between the Ba star ascending the RGB and its WD companion.

\begin{figure}[t]
\begin{center}
\includegraphics[width=0.49\textwidth]{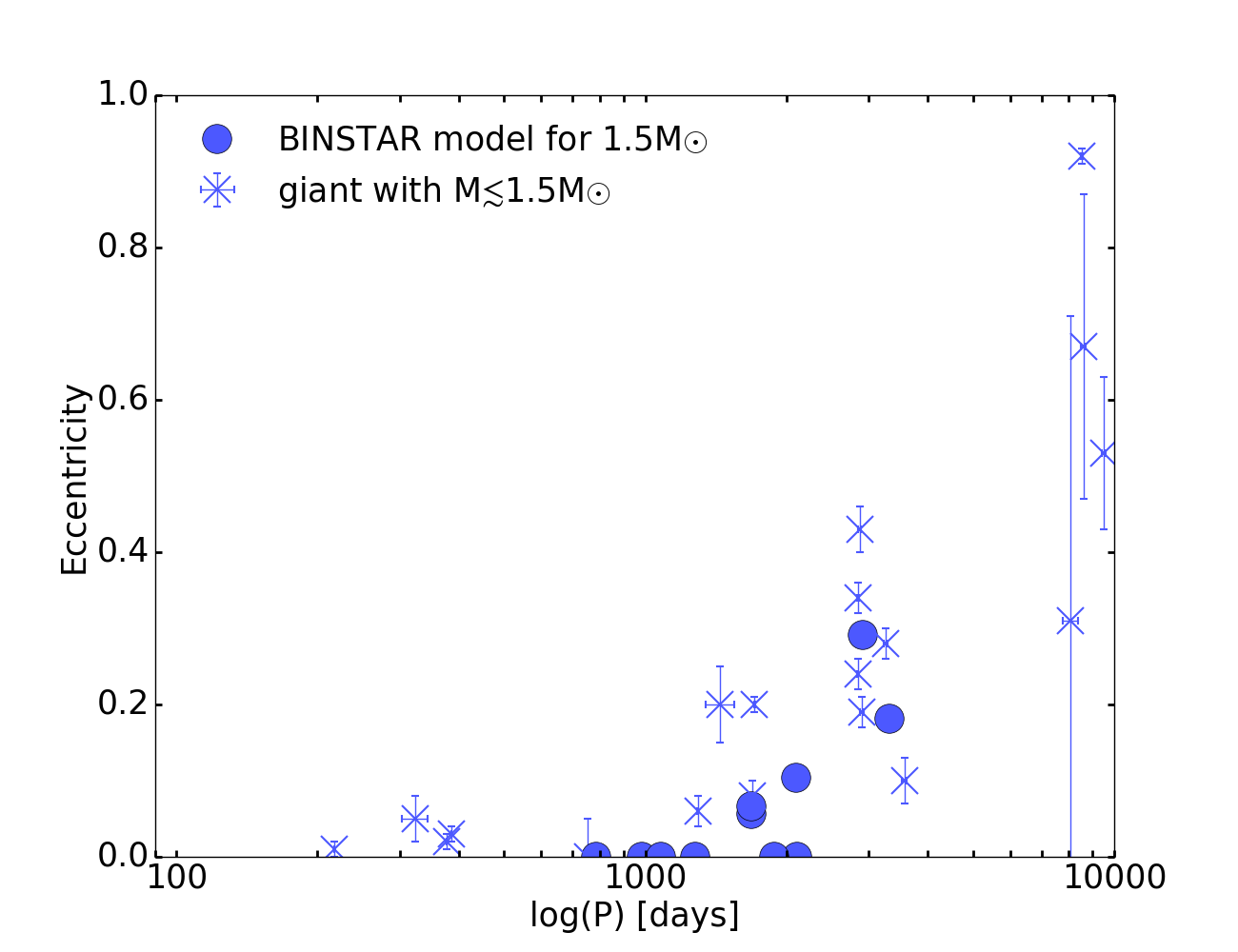}
\includegraphics[width=0.49\textwidth]{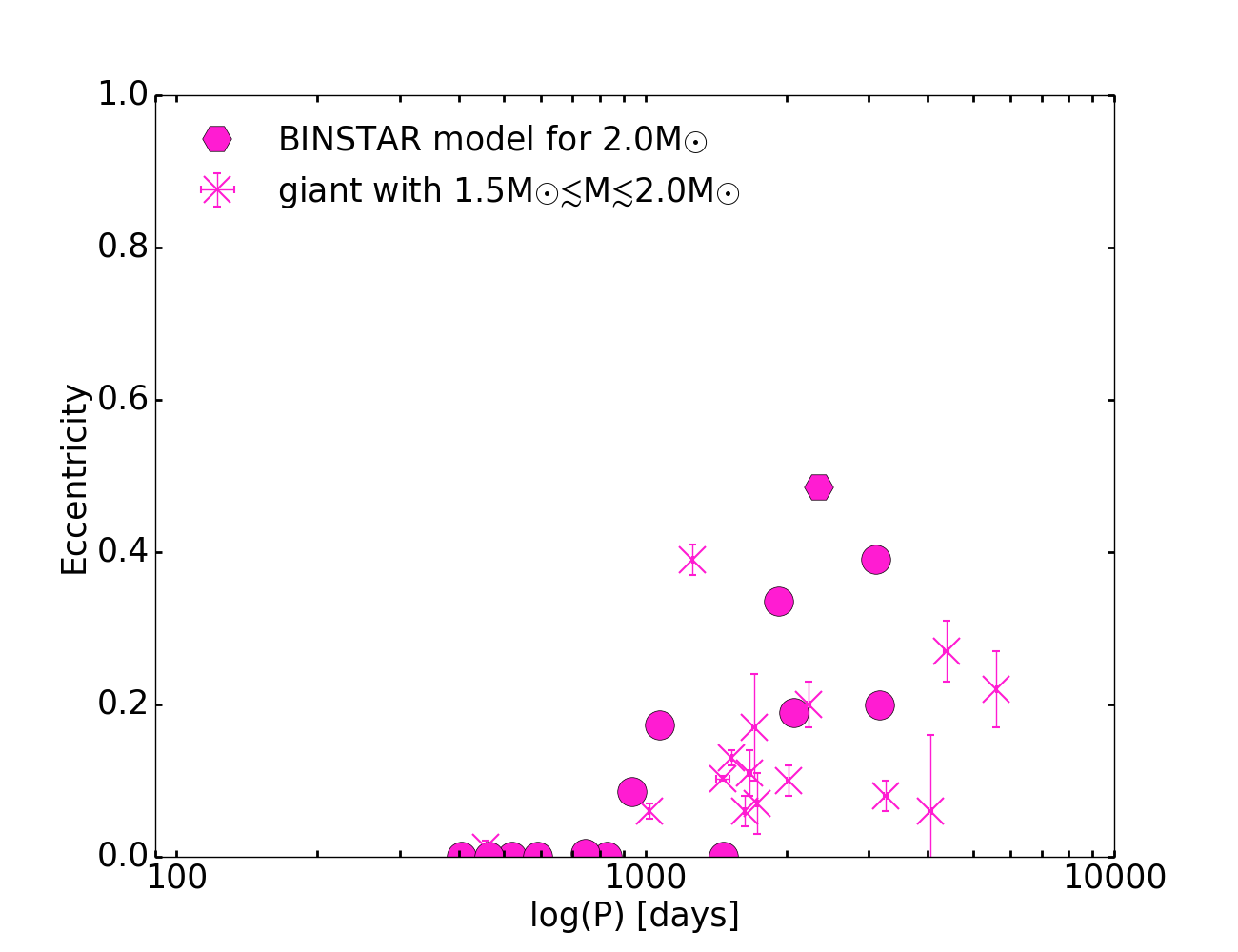}
\caption{\label{Fig:elogP} Observed and modelled orbits of Ba stars after the RGB phase.}
\end{center}
\end{figure}

\section{Methods}
In \cite[Escorza et al. (2017)]{Escorza17}, we presented a Hertzsprung-Russell diagram (HRD) of Ba stars using photometric stellar parameters and TGAS (Tycho-Gaia Astrometic Solution; \cite[Lindegren et al. 2016]{Lindegren16}) parallaxes. Our database includes a sample of 90 objects which have a fully covered binary orbit (\cite[Jorrissen et al. 2016]{Jorissen16} and Escorza et al. in prep). We redetermined the stellar parameters of these from
high-resolution spectra, and we used distances from \cite[Bailer-Jones et al. (2018)]{B-J18} to compute new, more accurate luminosities. Finally, we determined the individual masses of the Ba stars by comparing their location on the HRD with a new grid of STAREVOL \cite[(Siess 2006)]{Siess06} evolutionary models. 

We used observations (masses, periods and eccentricities) of the main sequence Ba stars as input parameters for a grid of standard binary evolutionary models computed with the BINSTAR code \cite[(Siess et al. 2013)]{Siess13}. The grid of initial parameters covered four initial masses: 1.5, 2.0, 2.5 and 3.0M$_{\odot}$; six orbital periods: 100, 300, 600, 1000, 2000 and 3000 days; and three eccentricities: 0.2, 0.4 and 0.6. Since our goal was to investigate the evolution of the orbits during the RGB phase of the Ba star, the secondary was chosen to be a cool WD, and we let the systems evolve until the onset of core He-burning. Then we compared the final orbital parameters with observations of Ba giant systems.

\section{Results and discussion}
Figure \ref{Fig:elogP} shows observed periods and eccentricities of Ba giants (crosses) and the final orbits of the models with 1.5 and 2.0M$_{\odot}$ that reached the core He-burning phase (circles). Systems in which the interaction made the star leave the RGB phase or systems that did not significantly interact (those with 2.5 and 3.0M$_{\odot}$) are not included in the figures.

Among the low-mass giants (M\,$\lesssim$\,1.5M$_{\odot}$), there are several systems with periods shorter than those predicted by the models (P\,$<$\,700\,days). Other systems with 1000\,day\,$\lesssim$ P $\lesssim$\,3000\,day are significantly more eccentric than the models of the corresponding mass and period. This indicates that additional interaction mechanisms are operating during the RGB phase, independently of the past interaction with the former AGB primary that led to the pollution of the present Ba star.

We now plan to test several mechanisms that might help us reproduce the observed orbits better, for example, a tidally enhanced mass-loss (e.g. \cite[Tout \& Eggleton 1988]{ToutEggleton88}) during the RGB phase or a reduction of the tidal efficiency (e.g. \cite[Nie et al. 2017]{Nie17}). Additionally, we plan to complement our observations with extrinsic S stars, which are thought to be the low-mass and more evolved counterparts of Ba stars, and might fall among the unpredicted orbits.

\end{document}